# Nonequilibrium Diagrammatic Technique for Nanoscale Devices

G. I. Zebrev

Department of Microelectronics, Moscow Engineering Physics Institute, Moscow, Russia

## ABSTRACTS

A general approach based on gauge invariance requirements has been developed for automatic construction of quantum kinetic equation in electron systems, far for equilibrium. Proposed theoretical scheme has high generality and automatism and capable to treat nonequilibrium effects of electron transport, quantum interference and energy dissipation. Dissipative and quantum-interference effects can be consequentially incorporated in the computational scheme through solution of dynamic Dyson equation for self-energies in the framework of conventional diagrammatic technique.

**Keywords:** nonequilibrium, nanoscale, Keldysh, kinetic equation, tunneling, open system, two-barrier structure.

## INTRODUCTION

Almost all promising versions of forthcoming transistor types have a three-part structure similar to conventional MOSFET: two massive contacts (leads) as "source" and "drain" separated by a "central region" (ballistic channel, quantum dot, molecular bridge etc.) which is electrostatically influenced by a control electrode ("gate"). The essential feature of the systems is their strongly nonequilibrium and open character. Electron devices are of use only when connected to a circuit, and to be so connected any device must have at least two terminals, contacts, or leads. As a consequence, every device is an open system with respect to electron flow.

Open systems drastically differs from closed ones. When a system is closed, it is isolated from its environment. Generally, closed systems are described by quantum mechanics laws. Environment has extremely large number of degrees of freedom. Therefore, in contrast to isolated systems the open systems demand quantum-theoretical (i.e. quantum-mechanical with infinitely large number of degrees of freedom) description.

The development of nanoscale semiconductor devices demands a clear and general description of nonequilibrium phenomena in microstructures. The efforts devoted to investigate electron transport through central part of such systems has sharply increased in recent years [1]. To describe electronic transport in nanoscale structures, in many cases we cannot resort to a classical Boltzmann equation and have to include the quantum-mechanical aspects of electronic transport.

On the other hand, in the time-reversible Schroedinger equation for an electron state, the state does not change its eigenenergy during its temporal evolution. Accordingly, this is a pure state description, which cannot treat electron-phonon and electron-electron interaction. Due to the statistical nature of kinetic processes, a definite conserved Hamiltonian for the Schroedinger equation cannot be specified and quantum device should be considered as a statistical system, characterized by the density matrix or non-equilibrium Green's function.

The objective of this work is to show how to derive kinetic equations for different electron systems in different environments based on very general ground of gauge invariance requirements.

## I. GENERAL BACKGROUND

Proposed approach is based on the following general ideas: (i) Preliminary averaging over environment or "bath" variables (phonon and/or impurities and/or "outer" electrons) (see, for example [2]); (ii) the gauge invariance requirements imposed on the open, nonequilibrium system. The former leads to irreversibility effects appearance in the system due to occurrence of non- hermiticity in Hamiltonian and non-unitarity in evolution operator.

Recall that introduction of imaginary negative addition $(-iV_1)$ to potential energy in quantum-mechanical Hamiltonian leads to appearance of non-zero negative term in right-hand-side of continuity equation

$$\frac{\partial \rho}{\partial t} + \nabla \mathbf{J} = -2\frac{V_1}{\hbar}\rho$$

where $\rho = \psi^*\psi$ is electron density and $\mathbf{J} = -(\hbar/m)\text{Im}(\psi^*\nabla\psi)$ is current density (see, for example, [3]). This simple form of imaginary potential implies that the one-electron motion decays exponentially with time and equations of motion become irreversible. The electron current in such way is not conserved due to the lack of hermiticity of Hamiltonian, which includes the complex potential [4]. We will show below that averaging over environment variables leads to appearance of collision integral in kinetic equation and in its turn to dissipation and irreversibility.

The key point of the proposed approach is a statement that kinetic equation can be derived in a consequential way based on very general gauge invariance requirements. As can be shown this heuristic approach provides exact conservation of an electron number and momentum and energy balance. An exact character of electron number conservation law implies a nonperturbative description of non-equilibrium kinetics.

In this section we consider for brevity only electron-phonon system in an applied external electric field with scalar and vector potential $A = (\varphi, \mathbf{A})$.

Generating functional of electron-phonon system ($U_A$ is evolution operator) of an open system can be written in path-integral form as

$$Z_A = tr(U_A^+ U_A) = \int D[\overline{\psi}]D[\psi]D[\phi]D[u]\exp(iS), \qquad (1)$$

where $U_A$ is evolution operator and $S$ is the action of electron-phonon system as a whole.

Effective action of electron-phonon system has a form of a convolution of matrices in the Keldysh space [5].

$$S = S_0 + S_{e-ph} + S_{ph} \qquad (2)$$

$$S_0 = \int dx\, \overline{\psi}_i \gamma_{ij}^k (\hat{G}_A^{-1})^k \psi_j = \int dx\{\overline{\psi}_1 \hat{G}_A^{-1} \psi_1 - \overline{\psi}_2 \hat{G}_A^{-1} \psi_2\}, \qquad (3)$$

where indices (1, 2) label forward and reverse Keldysh contour branches. Vertex tensor is defined here through third Pauli matrix $\tau_3$

$$\gamma_{ij}^k = (\tau_3)_{ik} \otimes \delta_{ij} \qquad (4)$$

Inverse Green's functions of non-interacting electrons are the same for two branches of Keldysh

$$(\hat{G}_A^{-1})^{(1)} = (\hat{G}_A^{-1})^{(2)} = i\partial_t + \mu - \varepsilon(\nabla - i(e/c)\mathbf{A}). \qquad (5)$$

Electron-phonon interaction is described be the action with free inverse phonon propagator $D^{-1}(x,x')$ and effective electron-phonon vertex $\Gamma_{ij}^k = g\gamma_{ij}^k$ ($g$ is electron-phonon coupling constant).

$$S_{e-ph} = \frac{1}{2}\int dx\, dx'\phi_i(x)[D^{-1}(x,x')]_{ij}\phi_j(x') + \int dx\, dx'\phi_i(x)\Gamma_{ij}^k\overline{\psi}_i(x)\psi_j(x'). \qquad (7)$$

### A. Averaging over thermostat variables

After functional integration over phonon fields (i.e. phase averaging over "bath" or thermostat variables) we have an effective "coarse-grained" action depending on only the electron variables

$$Z_A = \int D[\overline{\psi}]D[\psi]D[\phi]\exp(iS_0 + iS_{e-ph} + iS_{ph}) = \int D[\overline{\psi}]D[\psi]\exp(i\langle S \rangle), \qquad (8)$$

where $\langle S \rangle = S_0 + S_{eff}$ is "coarse-grained" action depending on only the electron variables and $S_{eff}$ is influence functional for electrons through interaction with phonons

$$S_{eff} = \frac{1}{2}\int dx\, dx'\overline{\psi}_n(x')\psi_l(x)\Gamma_{nl}^k(D(x,x'))^{kk'}\Gamma_{ij}^{k'}\overline{\psi}_i(x)\psi_j(x') \qquad (9)$$

### B. Gauge requirements

This functional (1) identically equals to unity in close equilibrium system due to unitarity of the evolution operator. This is not the case for non-equilibrium systems.

General requirement of gauge invariance imposed on the reduced generating functional can be written as

$$\langle \delta Z \rangle_C = \langle \delta Z(S_0) \rangle_C + \langle \delta Z(S_{eff}) \rangle_C = \left\langle \delta\overline{\psi}_i(x)(\tau_3)_{ij} \frac{\delta \langle S \rangle}{\delta \overline{\psi}_j(x)} + \frac{\delta \langle S \rangle}{\delta \psi_j(x)}(\tau_3)_{ji} \delta\psi_i(x) \right\rangle_C = 0 \quad (10)$$

where $\langle ... \rangle_C = \int d\overline{\psi} d\psi ... \exp(i\langle S \rangle)$ is the Keldysh contour ordered phase averaging. This is the key point of the approach. Let us consider symmetry of (1) under global $U(1)$-transformation

$$\delta\overline{\psi} = ie\Lambda\overline{\psi}, \quad \delta\psi = -ie\Lambda\psi, \quad (11)$$

where $e$ is electron charge, $\Lambda$ is infinitesimal dimensional parameter.

This invariance takes account of the exact and fundamental electric charge conservation law for non-equilibrium case. Resulting equations from the very outset have fully gauge invariant one-point and c-number form. This is similar to a situation in quantum mechanics, where wave function phase invariance leads to charge conservation.

## II. QUANTUM KINETIC EQUATION FOR UNIFORM SYSTEMS

### A. Hydrodynamic form of left hand side of kinetic equation

Standard left-hand-side of kinetic equation can be derived from free part $S_0$ of reduced action. Using conditions (10-11) and Wick's theorem for contour-ordered terms we can write in one-loop single-electron approximation

$$\langle \delta Z(S_0) \rangle_C = \int dx' \hat{L}(x',x)[G^K(x,x')] = \int dx'(-i)\left\{\delta(x-x')\left(\hat{G}_0^{-1}\right)G^K(x,x') - G^K(x,x')\left(\hat{G}_0^{-1}\right)\delta(x-x')\right\} \quad (12)$$

$$\hat{L}(x',x) = (-i)\left\{\delta(x-x')\left(\hat{G}_0^{-1}\right) - \left(\overline{\hat{G}}_0^{-1}\right)'\delta(x-x')\right\} = (-i)\delta(x-x')\left(\left(\hat{G}_0^{-1}\right) - \left(\hat{G}_0^{-1}\right)^*\right) \quad (13)$$

where $G^K = tr(G_{ij})$ is the Keldysh component of matrix Green's function. In further we will use a specific "triangle" Larkin-Ovchinnikov representation for Green's function (see Appendix) [6].

Let us consider uniform systems where semiclassical conditions $\overline{\varepsilon}\tau \gg 1$ и $\overline{p}\Omega^{1/d} \gg 1$, ($\tau$ and $\Omega^{1/d}$ are characteristic temporal and spatial scales of the system) allow to write the left-hand-side of kinetic equation (originating from "free" part of the action) in a standard "hydrodynamic" form. Semiclassical conditions give ground to examine two-point Green's function $G(x,x')$ as approximately translationally-invariant $G^K(x,x') \cong G^K(x-x')$, that allows to turn to momentum representation. Note that there is no need in a use of Wigner representation in this case. It is well-known that a use of mixed Wigner representation for two-point Green's functions leads to awkward gradient expansion (see, for example, [7]). Non-closed form of this Taylor expansion unsatisfactorily agrees with exact character of charge conservation. This problem arises only if we attempt to derive kinetic equations immediately from two-point Green's function with Dyson equations. Measured physical quantities such as electron densities and currents may be only expressed in terms of one-point functions (equal temporal and spatial coordinates). Gauge invariance requirement (10) immediately leads to one-point local description with no needs in gradient expansion and Wigner representation.

It can be shown that a result has the form of a kinetic equation for Keldysh component $G^K = tr(G_{ij})$ of Green's function $\hat{L}G_p^K = \Re_p^{e-p}$, where $\hat{L}$ is a hydrodynamic form of left-hand side of kinetic equation

$$\left(\partial_t + \mathbf{v}\nabla_\mathbf{r} + e\mathbf{E}_t\mathbf{v}\frac{\partial}{\partial \varepsilon} + \left(e\mathbf{E}_l + \frac{e}{c}(\mathbf{v}\times\mathbf{H})\right)\nabla_\mathbf{p}\right)G^K(x,p) \quad (14)$$

where $\mathbf{E}_l = -\nabla\varphi$, $\mathbf{E}_t = -\partial_t\mathbf{A}/c$ are longitudinal and transversal electric field in the Coulomb gauge).

### B. General form of electron-phonon collision integral

Collision integral stems in natural way from effective part of the action (7). Using the gauge requirements (10) and (11) it can be shown that an electron-phonon collision integral in general energy-momentum representation has a form

$$St_{e-ph} = \frac{1}{2}\int dq\left(\tau_1\underline{\Gamma}_q^k \underline{G}_p \underline{D}_q^{kk'} \underline{\widetilde{\Gamma}}_q^{k'} \underline{G}_{p+q} + \underline{G}_p \underline{\Gamma}_q^k \underline{G}_{p-q} \underline{D}_q^{kk'} \tau_1 \underline{\widetilde{\Gamma}}_q^{k'} - \underline{\Gamma}_q^k \tau_1 \underline{D}_q^{kk'} \underline{G}_{p-q} \underline{\widetilde{\Gamma}}_q^{k'} \underline{G}_p - \underline{G}_{p+q}\underline{\Gamma}_q^k \underline{D}_q^{kk'} \underline{G}_p \underline{\widetilde{\Gamma}}_q^{k'} \tau_1\right) \quad (15)$$

where shorthand notation $dq = d\varepsilon\, d^d\mathbf{q}/(2\pi\hbar)^d$ is introduced and $q=(\omega,\mathbf{q})$ and $p=(\varepsilon,\mathbf{p})$ are transfer energy-momentum vector.

This expression has a structure (*GE*+*GA* - *LA* - *LE*) "gain minus loss" due to phonon's emission and absorption processes. (*GE* (*LE*) and *GA*(*LA*) denote "ingoing" ("outgoing") terms due to phonon emission and absorption respectively). It is convenient to represent Eq.15 graphically using standard rules of diagram technique [8]. Here $p=(\varepsilon,\mathbf{p})$ is an energy-momentum vector. Solid (wavy) lines correspond to electron (phonon) propagator; full dot corresponds to the first Pauli matrix $\tau_1$ (see Appendix). Every line corresponds to electron (solid lines) and phonon (wavy lines) matrix Green's functions. Here we use a specific Larkin-Ovchinnikov matrix representation in Keldysh space function (see Appendix) where the full dot on diagrams corresponds to the first Pauli matrix $\tau_1$ and electron-phonon vertices are different for emission (marked with tilde) and absorption processes. Notice that, unlike in the equilibrium diagrams where directions of phonon lines are irrelevant, opposite directions interactions lines with necessity occurs for absorption and emission in non-equilibrium processes.

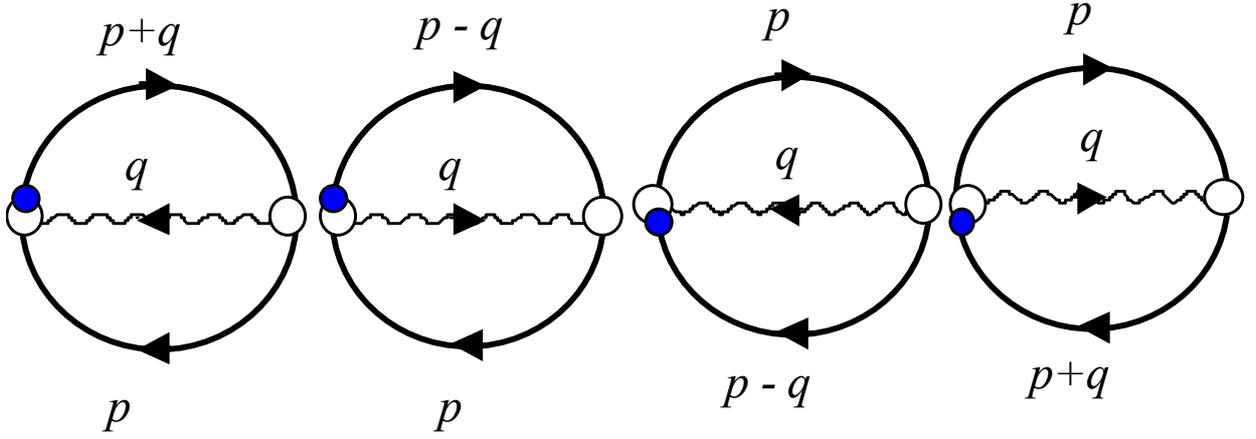

Fig.1 Diagram representation of electron-phonon collision integral

Direct computation of matrix convolution automatically yields electron-phonon collision integral which after integration over energy variable get a standard form

$$S_{e-ph} = 2\pi \int \frac{d^3\mathbf{q}}{(2\pi\hbar)^3} |g|^2$$
$$\{[(N_\mathbf{q}+1)n_{\mathbf{p+q}}(1-n_\mathbf{p}) - N_\mathbf{q}\, n_\mathbf{p}(1-n_{\mathbf{p+q}})]\delta(\varepsilon_{\mathbf{p+q}} - \varepsilon_\mathbf{p} - \omega_\mathbf{q}) +$$
$$+ [N_\mathbf{q}\, n_{\mathbf{p-q}}(1-n_\mathbf{p}) - (1+N_\mathbf{q})n_\mathbf{p}(1-n_{\mathbf{p-q}})]\delta(\varepsilon_{\mathbf{p-q}} - \varepsilon_\mathbf{p} - \omega_\mathbf{q})\} \quad (16)$$

To derive the momentum-and energy-balance equations one needs to consider the invariance of (11) with respect to infinitesimal spatial and temporal translation [9]. It can be shown that the right-hand side of a balance equation has a structure (*q*/2)(*GA* + *LA* - *GE* - *LE*),

## C. Symmetries of collision integral

Generally speaking all three non-zero components of matrix Green's function are independent but relate each with other under certain conditions. For example time-reversal symmetry is expressed through identity for advanced and retarded Green's functions

$$G^R(x,x') = \left(G^A(x',x)\right)^*, \quad (17)$$

Keldysh component is expressed in equilibrium as

$$G^K = (1-2n(\varepsilon))(G^R - G^A) \quad (18)$$

Symmetry of electron and phonon propagators can be written in a matrix form as follows

$$\left(\underline{G}_p\right)^+ = \tau_2 \underline{G}_p \tau_2; \qquad\qquad \left(\underline{D}_p\right)^+ = \tau_2 \underline{D}_p \tau_2 \qquad(19)$$

Electron-phonon vertices for absorption and emission due to Hermitian character of dynamic equation obey the relations

$$\left(\underline{\tilde{\Gamma}}_q^k\right)^+ = i\left(\tau_2 \underline{\Gamma}_q^{k'} \tau_2\right)(\tau_2)^{k'k} \qquad \left(\Gamma_q^k\right)^+ = i(\tau_2)^{k'k}\left(\tau_2 \underline{\tilde{\Gamma}}_q^{k'} \tau_2\right) \qquad(20)$$

Using these identities and trace invariance relative to transposition operation one can get
$$GE = LA^*, \qquad GA = LE^*. \qquad(21)$$
It can be shown that in equilibrium emission and absorption amplitudes are equal
$$GE = GA; \qquad LA = LE \qquad(22)$$
and collision integral can be simplified to a form 2Im(*GE+GA*).

## III. NONEQUILIBRIUM TECHNIQUE FOR TUNNEL STRUCTURES

### A. Two-barrier structure

Let us we have a system consisting of three parts (*a*, *b* and *c*) i.e., the central region and two contact reservoir.

Electron in the central region are coupled to those in the two leads with tunneling constants *T*, which are dependent in on the barrier height and width. After applying of external bias the state of left-lead electrons becomes different from that of the right-lead electrons and the total system including the quantum-well electrons is in nonequilibrium state.

Fig.3 schematically shows two-barrier structure with two massive contacts (*a*, *b*) and central quantum well (*c*) considered in this work.

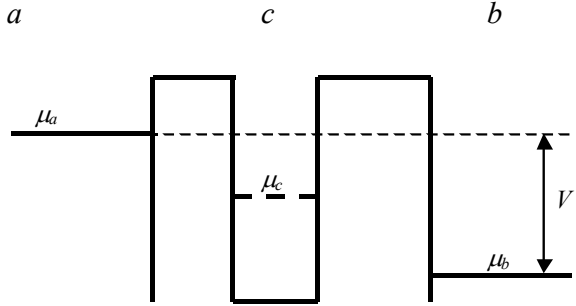

Fig.3 Idealized two-barrier structure

### B. The action for electrons in two-barrier system

We will not consider for simplicity space charge effects in the leads and single-electron charging effects in the central quantum well demanding self-consistent Poisson equation solution. Whole electron system can divided into three parts with the actions

$$S_a = \int dx_l dx_l' \,\bar{a}_i(x_l)\left(G_a^{-1}[x_l - x_l']\right)_{ij} a_j(x_l') \equiv \bar{a} \circ G_a^{-1} \circ a \qquad(23)$$

$$S_b = \int dx_r dx_r' \,\bar{b}_i(x_r)\left(G_b^{-1}[x_r - x_r']\right)_{ij} b_j(x_r') \equiv \bar{b} \circ G_b^{-1} \circ b ; \qquad(24)$$

$$S_c = \int dx_c dx_c' \,\bar{c}_i(x_l)\left(G_c^{-1}[x_c - x_c']\right)_{ij} c_j(x_c') \equiv \bar{c} \circ G_c^{-1} \circ c . \qquad(25)$$

We will assume here for brevity with loss of generality that central well is quite thin to consider tunneling through double-barrier structure to be coherent. This corresponds to zero collision integral due to dissipative processes, for example phonon emission or absorption.

Tunnel coupling between the central region and the leads can be written as

$$S_{tun} = \int dx_a dx_c\, \bar{a}_i(x_a)\gamma^k_{ij} T^k_{ac}(x_a - x_c) c_j(x_c) + \int dx_a dx_c\, \bar{c}_j(x_c)\gamma^k_{ji} T^k_{ca}(x_c - x_a) a_i(x_a) + \\ + \int dx_c dx_b\, \bar{c}_i(x_c)\gamma^k_{ij} T^k_{cb}(x_c - x_b) b_j(x_b) + \int dx_c dx_b\, \bar{b}_j(x_b)\gamma^k_{ji} T^k_{bc}(x_b - x_c) c_i(x_c) \quad (26)$$

or in symbolic brief form

$$S_{tun} = \langle \bar{a} \circ T_{ac} \circ c \rangle + \langle \bar{c} \circ T_{ca} \circ a \rangle + \langle \bar{c} \circ T_{cb} \circ b \rangle + \langle \bar{b} \circ T_{bc} \circ c \rangle. \quad (27)$$

The action for whole systems in such a way has a form

$$S_{abc} = S_a + S_b + S_c + S_{tun} = \\ = \bar{a} \circ G_a^{-1} \circ a + \bar{c} \circ G_c^{-1} \circ c + \bar{b} \circ G_b^{-1} \circ b + \bar{a} \circ T_{ac} \circ c + \bar{c} \circ T_{ca} \circ a + \bar{c} \circ T_{cb} \circ b + \bar{b} \circ T_{bc} \circ c \quad (28)$$

### C. Averaging over contact electron fields

To derive kinetic equation for electrons in the central region one needs to eliminate dependence on electrons in the leads which play role of "bath" variables. Performing exact Gaussian functional integration over electron fields ($a$ and $b$) we have

$$S_c = \bar{c} \circ \left(G_c^{-1} - T_{ca} \circ G_a \circ T_{ac} - T_{cb} \circ G_b \circ T_{bc}\right) \circ c \equiv \bar{c} \circ \mathbf{G}_c^{-1} \circ c. \quad (29)$$

The expression in brackets in (29) is nothing else than inverse Green's function for electrons in the central region renormalized due to tunneling exchange between the central region and contacts. This is expressed through standard Dyson equation

$$\mathbf{G}_c^{-1} = G_c^{-1} - \Sigma_a - \Sigma_b \equiv G_c^{-1} - T_{ca} \circ G_a \circ T_{ac} - T_{cb} \circ G_b \circ T_{bc}, \quad (30)$$

or in equivalent form

$$\mathbf{G}_c = G_c + G_c \circ (\Sigma_a + \Sigma_b) \circ \mathbf{G}_c. \quad (31)$$

Self-energy parts appear here due to tunnel coupling with the contacts

$$\Sigma_a + \Sigma_b \equiv T_{ca} \circ G_a \circ T_{ac} + T_{cb} \circ G_b \circ T_{bc}. \quad (32)$$

Tunnel amplitudes have time-reversal symmetry properties

$$T_{ac} = T_{ca}^* = T_a; \quad T_{bc} = T_{cb}^* = T_b. \quad (33)$$

Как и в предыдущем разделе, эти условия можно выразить в виде определенной симметрии туннельных вершин. В простейшем случае упругого туннелирования имеем

$$\widetilde{\gamma}^k\, T^*_{-\mathbf{q},-\mathbf{p}} = i\left(\tau_2 \underline{\gamma}^{k'} T_{\mathbf{p},\mathbf{q}} \tau_2\right)^{k'k} \qquad \underline{\gamma}^k\, T^*_{-\mathbf{q},-\mathbf{p}} = i\left(\tau_2 \underline{\gamma}^{k'}_q T_{\mathbf{q},\mathbf{p}} \tau_2\right)^{k'k}. \quad (34)$$

### D. Kinetic equation for electron system in the central region

Electron systems in the contact reservoirs ($a$, $b$) are assumed as usually to be in quasi-equilibrium with electrochemical potentials ($\mu_a$, $\mu_b$), and the Fermi occupancy numbers

$$n_a = n_F(\varepsilon - \mu_a), \quad n_b = n_F(\varepsilon - \mu_b). \quad (35)$$

Electron occupancies number in the electrodes are controlled by its potential biases.

Using gauge invariance requirement for the reduced action (29) kinetic equation for electron system in the central device part can be represented as follows

$$\hat{L}(G_c^K) = \sum_{k,k'} \sum_{\alpha=a,b} \left(\underline{G}_\alpha \circ \tau_1 \circ \widetilde{\underline{\gamma}}^k \circ \underline{G}_c \circ \underline{\gamma}^{k'} \circ \underline{T}^{kk'}_{ac} - \tau_1 \circ \underline{\gamma}^{k'} \circ \underline{G}_\alpha \circ \widetilde{\underline{\gamma}}^k \circ \underline{G}_c \circ \underline{T}^{k'k}_{ca}\right), \quad (36)$$

where $\hat{L}$ is total time derivative, $\underline{G}$ are matrix Green-Keldysh functions [2] for central device and contact parts (indices "$c$" and $\alpha=a,b$ respectively), $\underline{T}^{kk'}_{ac}$ are transfer tunnel elements, coupling different device's parts, $\underline{\gamma}^1_{ij} = \widetilde{\underline{\gamma}}^2_{ij} = \delta_{ij}/\sqrt{2};\ \underline{\gamma}^2_{ij} = \widetilde{\underline{\gamma}}^1_{ij} = (\tau_1)_{ij}/\sqrt{2}$ are bare vertex functions.

Corresponding diagrams are depicted in Fig.4.

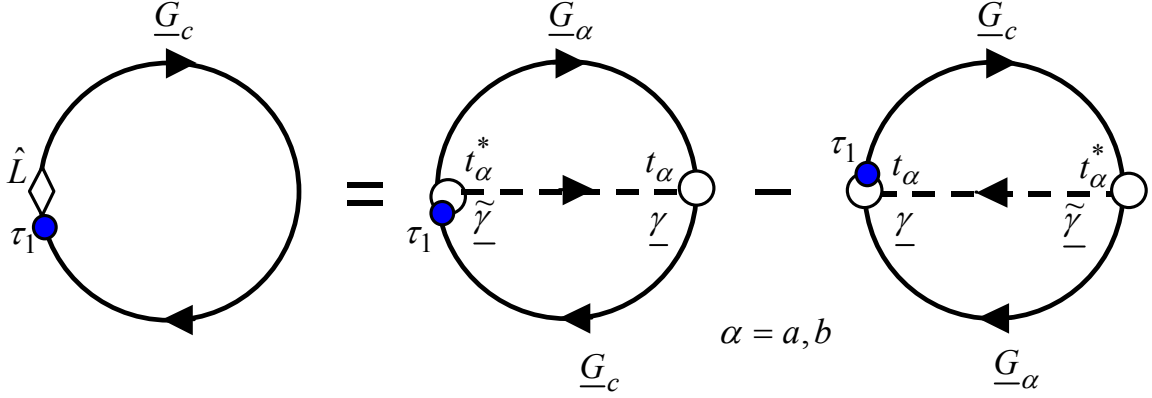

Fig.4. Diagrammatic representation of quantum kinetic equation Keldysh component of nonequilibrium matrix Green's function in the central dot of the system

For elastic tunneling

$$\left(T_\alpha\right)^{kk'} = \begin{pmatrix} 0 & 2i|t_\alpha|^2 \\ 0 & 0 \end{pmatrix} \tag{37}$$

$$|t(\mathbf{r}-\mathbf{r'})|^2 \propto \exp\left(-2\frac{\sqrt{2mU_b}}{\hbar}d\right)\delta_{\mathbf{p}_\perp \mathbf{p}_\perp'} \tag{38}$$

"Collision integral" in right-hand-side of (36) in lack of dissipation appears exclusively due the tunnel coupling of the central part with the contacts. It has additive form $St_c = St_{ac} + St_{bc}$, reflecting a possibility for electrons to be exchanged independently with both contacts.

Direct computation of 2×2 Green's function convolution traces yields

$$St_{ac} = 2i|t_a|^2\left[G_a^A G_c^K + G_a^K G_c^R\right] - 2i|t_a|^2\left[G_c^A G_a^K + G_c^K G_a^R\right] \tag{39}$$

$$St_{bc} = 2i|t_b|^2\left[G_b^A G_c^K + G_b^K G_c^R\right] - 2i|t_a|^2\left[G_c^A G_b^K + G_c^K G_b^R\right] \tag{40}$$

$$\partial_t G_c^K = 2i|t_a|^2(n_a - n_c)\left(G_a^R - G_a^A\right)\left(G_c^R - G_c^A\right) + 2i|t_b|^2(n_b - n_c)\left(G_b^R - G_b^A\right)\left(G_c^R - G_c^A\right) \tag{41}$$

### E. Steady-state regime

Steady-state condition corresponds to current equality through both barriers. This allows to compute the Keldysh component in the central part of the structure through the Green's function components in the contacts

$$G_c^K(\varepsilon,\mathbf{p}_c) = \frac{\sum_{\mathbf{p}_a}|t_a|^2 G_a^K(\varepsilon,\mathbf{p}_a) + \sum_{\mathbf{p}_b}|t_a|^2 G_b^K(\varepsilon,\mathbf{p}_b)}{\sum_{\mathbf{p}_a}|t_a|^2\left(G_a^R(\varepsilon,\mathbf{p}_a) - G_a^A(\varepsilon,\mathbf{p}_a)\right) + \sum_{\mathbf{p}_b}|t_b|^2\left(G_a^R(\varepsilon,\mathbf{p}_b) - G_b^A(\varepsilon,\mathbf{p}_b)\right)} \times \left(G_c^R(\varepsilon,\mathbf{p}_c) - G_c^A(\varepsilon,\mathbf{p}_c)\right) \tag{42}$$

Density of states in both contacts can be defined as ($\alpha = a, b$)

$$\nu_\alpha(\varepsilon) = \left(\frac{i}{2\pi}\right)\sum_{\mathbf{p}_\alpha}\left(G_\alpha^R(\varepsilon,\mathbf{p}_\alpha) - G_\alpha^A(\varepsilon,\mathbf{p}_\alpha)\right) \tag{43}$$

It is convenient to introduce tunneling probability from the dot to the contacts per a unit time $\gamma_\alpha$

$$\gamma_\alpha(\varepsilon) = \frac{1}{\hbar}\sum_{\mathbf{p}_\alpha}|t_\alpha|^2\left(G_\alpha^R(\varepsilon,\mathbf{p}_\alpha) - G_\alpha^A(\varepsilon,\mathbf{p}_\alpha)\right) = \frac{2\pi}{\hbar}|t_\alpha|^2\nu_\alpha(\varepsilon). \tag{44}$$

Using (42) and (43), one can get non-equilibrium Keldysh function in the c-region in a quasi-equilibrium form

$$G_c^K(\varepsilon,\mathbf{p}_c) = (1 - 2n_c(\varepsilon))\left(G_c^R(\varepsilon,\mathbf{p}_\alpha) - G_c^A(\varepsilon,\mathbf{p}_\alpha)\right) \tag{45}$$

where non-equilibrium distribution function $\mu_c$ for electrons in the central part of the structure is expressed through equilibrium Fermi distributions with electrochemical potentials $\mu_a$ and $\mu_b$ in the contacts

$$n_c(\varepsilon) = n_F(\varepsilon - \mu_c) = \frac{\gamma_a n_F(\varepsilon - \mu_a) + \gamma_b n_F(\varepsilon - \mu_b)}{\gamma_a + \gamma_b} \quad . \tag{46}$$

Steady-state condition leads in such manner to equation for non-equilibrium Fermi level position in the central part of the system.

Dissipative processes can be incorporated in the technique by means of standard diagram techniques.

## IV. CONCLUSIONS

We have developed a general technique for derivation of quantum kinetic equation based on very general requirements. It was shown that gauge invariance requirements imposed on a system lead to an exact form of Boltzmann-type quantum kinetic equation with generalized "collision integral" describing dissipative effects and/or charge exchange with reservoirs ("source" and "drain"). It has been shown that structure of kinetic equation can be exactly expressed in diagrammatic form.

Proposed technique has high generality and provides a convenient tool for description of nonequilibrium processes in different non-equilibrium electron systems in various approximations. The effects of screening, interference between various types of scattering, quantum coherence phenomena can be included in this approach according perturbation theory by means of a standard diagrammatic consideration.

## APPENDIX

We use here the Keldysh diagram technique in the specific representation [6,7], where $\tau_1, \tau_2, \tau_3$ are the Pauli matrices, where single-electron and phonon Green's functions have following matrix structure where $G^K = tr(\tau_1 \underline{G})$:

$$\underline{G}_{ab}(x, x') = -i \langle \psi_a(x) \overline{\psi}_b(x') \rangle = \begin{pmatrix} G^R(x, x') & G^K(x, x') \\ 0 & G^A(x, x') \end{pmatrix} \tag{A1}$$

where retarded (R) and advanced (A) Green's functions describe dynamical effects,

$$G_p^R = \frac{1}{\varepsilon - \varepsilon(\vec{p}) + i/\tau}; \qquad G_p^A = \left(G_p^R\right)^* \tag{A2}$$

and Keldysh component

$$G^K = (1 - 2n(\varepsilon))(G^R - G^A) \tag{A3}$$

Phonon Green's function can be written in Larkin-Ovchinnikov representation in a similar triangle form

$$\underline{D} = \begin{pmatrix} D^R & D^K \\ 0 & D^A \end{pmatrix} \tag{A3}$$

$$D_q^R = \frac{1}{\omega - \omega(q) + i\delta} - \frac{1}{\omega + \omega(q) + i\delta}; \qquad D_q^A = \left(D_q^R\right)^* \tag{A6}$$

and the Keldysh component $G^K$ is responded to statistical effects and expressed in equilibrium (or near equilibrium) via Fermi-Dirac ($n(\varepsilon)$) or Bose ($N(\varepsilon)$) distribution functions.

$$G^K = (1 - 2n(\varepsilon))(G^R - G^A), \quad D^K = (1 + 2N(\varepsilon))(D^R - D^A) \tag{A7}$$

Bare vertices for emission (with a tilde) and absorption processes have a structure

$$\underline{\gamma}_{ij}^1 = \underline{\widetilde{\gamma}}_{ij}^2 = \delta_{ij}/\sqrt{2}, \; \underline{\gamma}_{ij}^2 = \underline{\widetilde{\gamma}}_{ij}^1 = (\tau_1)_{ij}/\sqrt{2} \tag{A8}$$